\documentclass[twocolumn,preprintnumbers,amsmath,amssymb]{revtex4}
\usepackage{graphicx,psfrag}
\begin{document}
\preprint{CLNS~06/1979, FERMILAB-PUB-06-364-T}%, MZ-TH/06-nn}

\title{\boldmath%
Analysis of ${\rm Br}(\bar B\to X_s\gamma)$ at NNLO with a Cut on Photon Energy
\unboldmath}

\author{Thomas Becher$^a$}
\author{Matthias Neubert$^{b,c}$}

\affiliation{$^a$\,Fermi National Accelerator Laboratory, P.O. Box 500, 
Batavia, IL 60510, U.S.A.\\
$^b$\,Institute for High-Energy Phenomenology, Laboratory for 
Elementary-Particle Physics, Cornell University, Ithaca, NY 14853, U.S.A.\\
$^c$\,Institut f\"ur Physik (ThEP), Johannes 
Gutenberg-Universit\"at, D--55099 Mainz, Germany}

\date{\today}

\begin{abstract}
By combining a recent estimate of the total $\bar B\to X_s\gamma$ branching 
fraction at $O(\alpha_s^2)$ with a detailed analysis of the effects of a cut 
$E_\gamma\ge 1.6$\,GeV on photon energy, a prediction for the partial 
$\bar B\to X_s\gamma$ branching fraction at next-to-next-to-leading order in 
renormalization-group improved perturbation theory is obtained, in which 
contributions from all relevant scales are properly factorized. The result 
$\mbox{Br}(\bar B\to X_s\gamma)=(2.98\pm 0.26)\cdot 10^{-4}$ is about 
$1.4\sigma$ lower than the experimental world average. This opens a window for 
significant New Physics contributions in rare radiative $B$ decays.
\end{abstract}

\maketitle

\section{Introduction}

The inclusive decay $\bar B\to X_s\gamma$ is an important example of a 
flavor-changing neutral current process, which has been used to test the 
flavor sector of the Standard Model. Many groups have worked on improving the 
theoretical analysis of this process so as to keep pace with refinements in 
the measurements of its branching fraction. The effective weak Hamiltonian at
next-to-next-to-leading order (NNLO) has been obtained by calculating 
multi-loop matching coefficients and anomalous dimensions 
\cite{Misiak:2004ew,Gorbahn:2004my,Gorbahn:2005sa,Czakon:2006notyet}. While 
the fermionic NNLO corrections to the $b\to s\gamma$ matrix elements 
have been known for some time \cite{Bieri:2003ue}, complete NNLO corrections 
are presently only available for the electro-magnetic dipole operator
\cite{Blokland:2005uk, Asatrian:2006ph}. However, an approximate result for the NNLO 
charm-penguin contributions has just been published \cite{Misiak:2006ab}. 
Combining these ingredients, a first estimate of the $\bar B\to X_s\gamma$ 
branching ratio at NNLO has been presented in \cite{Misiak:2006zs}.

A complication in the analysis arises from the fact that measurements of the 
$\bar B\to X_s\gamma$ branching fraction impose stringent cuts on photon energy 
(defined in the $B$-meson rest frame), $E_\gamma>E_0$, with $E_0$ in the range 
between 1.8 to 2.0\,GeV. The standard treatment is to extrapolate different 
measurements to a common reference point $E_0=1.6$\,GeV using 
phenomenological models \cite{Buchmuller:2005zv}. In that way, 
the experimental world average 
$\mbox{Br}(\bar B\to X_s\gamma)=(3.55\pm 0.24_{\,-0.10}^{\,+0.09}\pm 0.03)\cdot%
10^{-4}$ has been derived \cite{HFAG:2006bi}. The first 
error is statistical, the second one systematical, the third one is due to the 
extrapolation from high $E_0$ to the reference value, and the last error 
accounts for the subtraction of $\bar B\to X_d\gamma$ background. A
theoretical result for the branching ratio with a cut at $E_0=1.6$\,GeV has 
been derived in \cite{Misiak:2006zs} using two-loop calculations of the 
photon-energy spectrum in 
fixed-order perturbation theory \cite{Melnikov:2005bx,Asatrian:2006sm}. It has 
been argued that the extrapolation from the total to 
the partial branching 
fraction does not introduce additional theoretical 
uncertainties. This assertion is questionable because of the dynamical 
relevance of a soft scale $\Delta=m_b-2E_0\approx 1.4$\,GeV, whose value is 
significantly lower than the $b$-quark mass.
 
Accounting for the photon-energy cut properly requires to disentangle 
contributions associated with the hard scale 
$\mu_h\sim m_b$, the soft scale $\mu_0\sim\Delta$, and an intermediate scale 
$\mu_i\sim\sqrt{m_b\Delta}$ set by the typical final-state hadronic invariant 
mass. When the cut value $E_0$ is chosen sufficiently low, $\Delta$ 
becomes a short-distance scale, and renormalization-group (RG) improved 
perturbation theory can be employed to calculate the effects of the 
photon-energy cut using a multi-scale operator product expansion 
\cite{Neubert:2004dd}. We have recently performed a systematic 
analysis of these effects at NNLO. Two-loop corrections at the soft scale were 
calculated in \cite{Becher:2005pd}, while those at the intermediate scale were 
computed in \cite{Becher:2006qw}. Here, the analysis is completed by extracting 
the two-loop hard matching corrections from a comparison with 
fixed-order calculations of the photon spectrum.

Using this method, we compute the fraction of all $\bar B\to X_s\gamma$ 
events with $E_\gamma\ge 1.6$\,GeV with a perturbative precision 
of 5\%. At this level of accuracy several other, nonperturbative effects need 
to be evaluated carefully. The event fraction receives hadronic power 
corrections $\sim(\Lambda_{\rm QCD}/\Delta)^n$ governed by 
$B$-meson matrix elements of local operators. The leading correction ($n=2$) 
is known and turns out to be small, but terms with $n\ge 3$ are presently 
unknown. Recently, a new class of enhanced $\Lambda_{\rm QCD}/m_b$ corrections to 
the $\bar B\to X_s\gamma$ decay rate has been identified, which involve 
matrix elements of nonlocal operators 
\cite{Lee:2006wn}. A model analysis using the vacuum insertion approximation 
indicates that these corrections 
affect the total decay rate at the level of a few percent. 

Combining our result for the event fraction with the prediction for the total
branching fraction from \cite{Misiak:2006ab,Misiak:2006zs}, we obtain
\begin{equation}\label{upshot}
   \mbox{Br}(\bar B\to X_s\gamma) = (2.98\pm 0.26)\cdot 10^{-4}
\end{equation}
for $E_0=1.6$\,GeV, where we have added in quadrature the uncertainties from 
higher-order perturbative effects (${}_{-6}^{+4}$\%), hadronic power corrections (5\%), 
parametric dependencies (4\%), and the interpolation in the charm-quark mass 
(3\%). Two-loop perturbative corrections at the intermediate and soft scales 
significantly lower the branching fraction 
with regard to the fixed-order result given in 
\cite{Misiak:2006ab}, and they 
increase the theoretical uncertainty.

\section{Scale Separation and Resummation}

At leading power in $\Lambda_{\rm QCD}/m_b$, the $\bar B\to X_s\gamma$ decay 
rate with a cut on photon energy obeys the factorization formula 
\cite{Korchemsky:1994jb}
\begin{eqnarray}\label{GammaOPE}
   \Gamma(E_0) &=& \frac{G_F^2\alpha}{32\pi^4}\,|V_{tb} V_{ts}^*|^2\,
    \overline{m}_b^2(\mu) |H_\gamma(\mu)|^2\!\!\int_0^\Delta\!\!\!dp_+(m_b-p_+)^3
    \nonumber\\
   &&\times \int_0^{p_+}\!d\omega\,m_b\,J(m_b(p_+ -\omega),\mu)\,S(\omega,\mu) \,,
\end{eqnarray}
where $p_+=m_b-2E_\gamma$ and $\Delta=m_b-2E_0$. The function $H_\gamma$ contains 
hard quantum corrections, the jet function $J$ describes the physics of the 
hadronic final-state jet, and the shape function $S$ parameterizes bound-state 
effects inside the $B$-meson \cite{Neubert:1993ch}. In the region of interest 
to our analysis, in which the quantity $\Delta\approx 1.4$\,GeV can be treated 
as a perturbative scale, the double convolution integral in (\ref{GammaOPE}) 
can be evaluated using short-distance methods \cite{Neubert:2004dd}. In order 
to isolate the effect of the photon cut, we focus on the fraction of all 
$\bar B\to X_s\gamma$ events that pass the cut $E_\gamma\ge E_0$, defined as 
$F(E_0)=\Gamma(E_0)/\Gamma(0)$. Since the total decay rate can be computed in 
fixed-order perturbation theory at the hard scale $\mu_h$, the event fraction 
obeys a factorization formula of the same form as (\ref{GammaOPE}), but with a 
different hard function $h$ in place of $|H_\gamma|^2$. 

Contributions associated with different mass scales can be separated from each 
other by evolving the various objects in the factorization formula away from 
a common renormalization scale $\mu$ to different ``matching scales'', where 
they can be calculated reliably using fixed-order perturbation theory. In this 
process, single and double logarithms of ratios of the different scales are 
resummed to all orders. The matching scales should be taken close to the 
default values $\mu_h=m_b$ for $H_\gamma$, $\mu_i=\sqrt{m_b\Delta}$ for $J$, 
and $\mu_0=\Delta$ for $S$. An elegant expression describing the RG evolution 
of the jet function was derived in \cite{Becher:2006nr}. It involves an 
associated jet function $\widetilde j$, which is related to $J$ 
by a Laplace transform. The same technique can be applied to describe the 
evolution of the shape function in terms of an associated soft function 
$\widetilde s$. Inserting these results into (\ref{GammaOPE}), the integrations 
over $p_+$ and $\omega$ can be performed, leading to
\begin{eqnarray}\label{master}
   F(E_0) &=& U(\mu_h,\mu_i,\mu_0; \mu) \nonumber\\
   &&\hspace{-1.3cm}\times \left( \frac{m_b}{\mu_h} \right)^{-2a_\Gamma(\mu_h,\mu)} 
    \left( \frac{m_b\Delta}{\mu_i^2} \right)^{2a_\Gamma(\mu_i,\mu)} 
    \left( \frac{\Delta}{\mu_0} \right)^{-2a_\Gamma(\mu_0,\mu)} \nonumber\\
   &&\hspace{-1.3cm}\times\, h\bigg( \frac{m_b}{\mu_h} \bigg)\,
    \tilde j\bigg( \ln\frac{m_b\Delta}{\mu_i^2} + \partial_\eta \bigg)\,
    \tilde s\bigg( \ln\frac{\Delta}{\mu_0} + \partial_\eta \bigg)\,
    \frac{e^{-\gamma_E\eta}}{\Gamma(1+\eta)} \nonumber\\
   &&\hspace{-1.3cm}\times \left[  p_3\Big( \frac{\Delta}{m_b} \Big)
    - \frac{\eta(1-\eta)}{6}\,\frac{\mu_\pi^2}{\Delta^2} 
    + \dots \right] + \delta F(E_0) ,
\end{eqnarray}
where $\eta=2a_\Gamma(\mu_i,\mu_0)>0$. Despite appearance, this result is 
independent of the choice of $\mu$ and of the three matching scales $\mu_h$, 
$\mu_i$, and $\mu_0$. For the special case $\mu=\mu_i$, relation 
(\ref{master}) coincides with a formula derived previously in 
\cite{Neubert:2005nt}. The terms in the first two lines arise from the RG 
resummation of single and double logarithms. The precise form of the evolution 
factor $U$ will be given in \cite{inprep}. The exponent 
\begin{equation}
   a_\Gamma(\mu_1,\mu_2) = \int_{\mu_1}^{\mu_2}\!\frac{d\mu}{\mu}\,
   \Gamma_{\rm cusp}(\alpha_s(\mu))
\end{equation}
is an integral over the cusp anomalous dimension. The functions $h$, 
$\widetilde j$, and $\widetilde s$ in the third line contain the matching 
corrections at the hard, intermediate, and soft scales, respectively. The 
two-loop expression for the jet function $\widetilde j$ has been obtained in 
\cite{Becher:2006qw}, while the two-loop result for the soft function 
$\widetilde s$ can be deduced from \cite{Becher:2005pd}. In the argument of 
these functions, $\partial_\eta$ means a derivative with respect to the 
quantity $\eta$. The hard function can be derived by matching (\ref{master}) 
with the fixed-order expression for the photon spectrum derived in 
\cite{Melnikov:2005bx,Asatrian:2006sm}. At the default scale $\mu_h=m_b$, we 
find
\begin{eqnarray}
   h(1) &=& 1 + \frac{C_F\alpha_s}{4\pi} 
    \bigg[ - \frac{52}{3} + \frac{7\pi^2}{6}
    - \sum_{i\le j}\,\mbox{Re}\frac{C_i^*C_j}{|C_7|^2}\,\hat f_{ij}(1) \bigg]
    \nonumber\\
   &&\mbox{}+ C_F \left( \frac{\alpha_s}{4\pi} \right)^2
    \left[ C_F H_F + C_A H_A + T_F n_f H_f \right] ,
\end{eqnarray}
where the term with $i=j=7$ is to be excluded from the sum. The one-loop 
result agrees with \cite{Neubert:2004dd}, and the two-loop coefficients are 
given by
\begin{eqnarray}
   H_F &=& \frac{2297}{24} - \frac{229\pi^2}{18} + \frac{89\pi^4}{360}
    - 22\zeta_3 + 16 S_{\rm \!a} \,, \nonumber\\
   H_A &=& - \frac{50521}{648} + \frac{1259\pi^2}{108} - \frac{5\pi^4}{18}
    + \frac{313}{9}\,\zeta_3 + 16 S_{\rm \!na} \,, \nonumber\\
   H_f &=& \frac{9365}{162} - \frac{19\pi^2}{27} - \frac{92}{9}\,\zeta_3 \,.
\end{eqnarray}
The constants $S_{\rm \!a}\approx 1.216$ and $S_{\rm \!na}\approx -4.795$ have 
been obtained by numerical integration in \cite{Melnikov:2005bx}. The complete 
expression for $h(m_b/\mu_h)$ 
including scale dependence will be given in \cite{inprep}.

In the last line of (\ref{master}), the hadronic quantity $\mu_\pi^2$ 
parameterizes the $B$-meson matrix element of the kinetic operator in 
heavy-quark effective theory. The ellipses represent unknown hadronic power 
corrections of order $(\Lambda_{\rm QCD}/\Delta)^3$. The polynomial
\begin{equation}
   p_3(\delta)
   = 1 - \frac{3\delta\eta }{1+\eta } + \frac{3\delta^2\eta}{2+\eta}
   - \frac{\delta^3\eta}{3+\eta}
\end{equation}
derives from the $(m_b-p_+)^3$ prefactor in (\ref{GammaOPE}). The term 
$\delta F(E_0)$ in (\ref{master}) collects terms that are power suppressed in 
the ratio $\delta=\Delta/m_b\approx 0.3$. These corrections are known in 
fixed-order perturbation theory, but no resummation is available. We express 
them as a perturbative series in powers of $\alpha_s(\mu)$. The one-loop 
result for $\delta F$ has been given in \cite{Neubert:2004dd}, while the 
dominant two-loop contribution has recently been calculated in 
\cite{Melnikov:2005bx,Asatrian:2006sm}. For subleading contributions from 
other operators in the effective Hamiltonian only the fermionic two-loop 
corrections are known \cite{Ligeti:1999ea}; they have a negligible impact on 
our results.

\section{Results and Conclusions}

In (\ref{master}), the parameters $m_b$ and $\mu_\pi^2$ are defined in the 
on-shell scheme. To improve the perturbative behavior, one should eliminate 
them in favor of appropriately defined short-distance quantities. 
We use the ``shape-function scheme'' proposed in \cite{Bosch:2004th}, 
in which low-scale subtracted heavy-quark parameters are defined via the 
moments of the renormalized $B$-meson shape function, regularized with a hard
cutoff $\mu_f$. The two-loop relations between the shape-function and 
pole-scheme parameters have been derived in \cite{Neubert:2004sp}. We adopt 
the conventional choice $\mu_f=1.5$\,GeV and use the values 
$m_b=(4.61\pm 0.06)$\,GeV and $\mu_\pi^2=(0.15\pm 0.07)$\,GeV$^2$ extracted 
from a fit to moments of inclusive $B$-decay spectra \cite{Neubert:2005nt}. 

\begin{figure}
\begin{center}
\psfrag{0.85}{\scriptsize $0.85$}
\psfrag{0.8}{\scriptsize  $0.8$}
\psfrag{0.95}{\scriptsize $0.95$}
\psfrag{0.9}{\scriptsize $0.9$}
\psfrag{1.}{\scriptsize $1$}
\psfrag{x}[]{\phantom{abc}\raisebox{-0.25cm}{$\mu_x/\mu_x^{\rm default}$}}
\psfrag{a}[l]{\scriptsize $1$}
\psfrag{d}[]{\scriptsize ${1}/{\sqrt{2}}$}
\psfrag{e}[]{}
\psfrag{c}[]{\scriptsize $\sqrt{2}$}
\psfrag{b}[]{}
\includegraphics[width=\columnwidth,height=0.55\columnwidth]{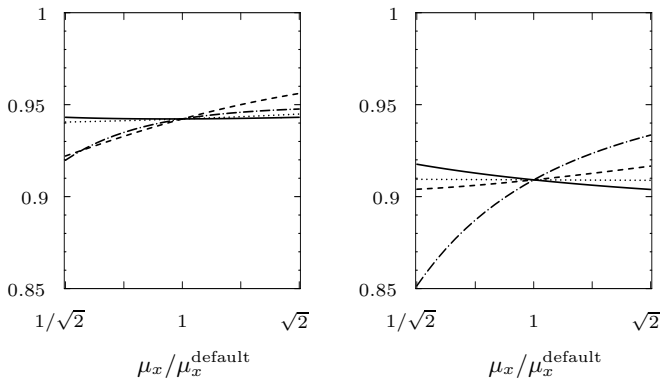}
\vspace{-0.7cm}
\end{center}
\caption{\label{fig:T}%
Scale dependence of the ratio $T$. The curves show the effect of 
varying the hard scale $\mu_h$ (dotted), intermediate scale $\mu_i$ (dashed), 
soft scale $\mu_0$ (dash-dotted), and the reference scale $\mu$ (solid) by a 
factor of $\sqrt{2}$ about their default values. In the left plot the 
resummation is performed for the leading-power terms only, while in the right 
plot the $p_3$ term is also included (see text for further explanation).}
\end{figure}

Ref.~\cite{Misiak:2006ab} does not provide a result for the total 
$\bar B\to X_s\gamma$ branching fraction. The most inclusive quantity 
considered is the partial branching fraction defined with a mild cut at 
$E_0=1$\,GeV. In order to combine this prediction with our RG-improved result 
for the event fraction, we define the ratio 
$T=F(1.6\,\mbox{GeV})/F(1.0\,\mbox{GeV})$. 
We evaluate the various matching scales at their default values, namely 
$\mu_h=m_b$, $\mu_0=\Delta$, and $\mu_i=\mu=\sqrt{m_b\Delta}$, where the 
values of $\Delta$ are different in the numerator and denominator. 
In order to study the residual 
scale dependence, we then vary each of the four scales by a factor of 
$\sqrt2$, correlated between numerator and denominator. While the ratio $T$ is 
formally independent of these scales, the residual dependence of the truncated 
perturbative expression can be taken as an estimate of higher-order effects. 
The results are depicted in Figure~\ref{fig:T}. Not surprisingly, the dominant 
effect arises from varying the lowest scale, $\mu_0\sim\Delta$, while the 
scale variations at the intermediate and high scales have a lesser impact.

The term proportional to $p_3(\Delta/m_b)$ in (\ref{master}) includes a subset 
of power corrections associated with a phase-space factor. While it was 
possible to perform the scale separation for these terms, treating them in a 
different way than the remaining power corrections in $\delta F(E_0)$ is 
somewhat arbitrary. The right plot in the figure refers to the form of the 
factorization formula shown in (\ref{master}), while the left plot 
corresponds to expanding out $(p_3-1)$ in fixed-order perturbation theory and 
including it in the $\delta F$ term. While the stability with respect to 
variations of the soft scale is better in this case, the perturbative 
corrections turn out to be smaller when performing the resummation for the 
$p_3$ term. The shift in central value between the two schemes is about 3\%, 
which is inside the error bar. This effect hints at the importance of RG resummation
for the power corrections.

To quote our final result we take the average of the two schemes and assign an 
asymmetric error reflecting the scale variation. This yields
\begin{equation}
   T = 0.93_{\,-0.05}^{\,+0.03}{}_{\rm pert}\pm 0.02_{\rm hadr}\pm 0.02_{\rm pars} \,.
\end{equation}
The event fraction $F(E_0)$ receives hadronic power corrections not suppressed 
by inverse powers of $m_b$, but only by powers of the soft scale $\Delta$ 
\cite{Neubert:2004dd}. 
These corrections are governed by $B$-meson matrix elements of local 
operators. The leading effect proportional to $\mu_\pi^2$ in (\ref{master}) is 
small mainly due to the smallness of its coefficient $\eta(1-\eta)/6$. 
Generically, we expect subleading corrections to scale like 
$\eta\,(\Lambda_{\rm QCD}/\Delta)^3$, for which we assign a 2\% uncertainty. 
The main uncertainties from parameter variations are $\mp 0.9\%$ for 
$\alpha_s(m_Z)=0.1189\pm 0.0020$, $\pm 0.4\%$ for $m_c/m_b=0.26\pm 0.03$, and 
$\pm 0.1\%$ for $m_b=(4.61\pm 0.06)$\,GeV. We also include a variation of 
$\pm 1.2\%$ 
due to the fact that the three-loop anomalous dimension of the shape function 
is yet unknown \cite{inprep}.
Our value for the ratio $T$ is lower than the estimate 
$T=0.963$ obtained in fixed-order perturbation theory 
\cite{Misiak:2006zs}. Moreover, we find that there is a significant theoretical 
uncertainty inherent in the calculation of $T$.

In order to complete the analysis we need as input the theoretical result for 
the $\bar B\to X_s\gamma$ branching fraction with $E_0=1.0$\,GeV, which we 
take from the fixed-order NNLO calculation of 
\cite{Misiak:2006ab,Misiak:2006zs}. These authors find
$\mbox{Br}(\bar B\to X_s\gamma)=(3.27\pm 0.23)\cdot 10^{-4}$ for $E_0=1$\,GeV, 
where the error has been obtained by adding in quadrature the uncertainties 
from higher-order perturbative corrections (3\%), nonperturbative effects 
(5\%), parameter dependencies (3\%), and the interpolation in the charm-quark 
mass employed in the NNLO estimate of charm-penguin loop graphs (3\%). The 
$\bar B\to X_s\gamma$ branching fraction receives incalculable power 
corrections starting at order $\Lambda_{\rm QCD}/m_b$, which cannot be 
described using the operator product expansion \cite{Voloshin:1996gw}. While 
we disagree with \cite{Misiak:2006ab} on the statement that these effects are 
suppressed by a power of $\alpha_s(m_b)$, we nevertheless believe that the 
corresponding uncertainty should not be larger than 5\%. Recently, a new class 
of enhanced power corrections has been identified \cite{Lee:2006wn}. At tree level 
their effects are parameterized in terms of $B$-meson matrix elements of 
nonlocal four-quark operators. Using the vacuum insertion approximation, a 
reduction of the total branching fraction between 0.3\% and 3\% has been found.
Accounting for this effect lowers the central value from 3.27 to 3.22.

Our final prediction for the $\bar B\to X_s\gamma$ branching fraction with 
$E_0=1.6$\,GeV is
\begin{eqnarray}\label{final}
   &&\hspace{-0.4cm} \mbox{Br}(\bar B\to X_s\gamma) \\
   &=& (2.98_{\,-0.17}^{\,+0.13}{}_{\rm pert}\pm 0.16_{\rm hadr}\pm 0.11_{\rm pars}
        \pm 0.09_{m_c})\cdot 10^{-4} \,, \nonumber
\end{eqnarray}
where we have combined errors of the same type in quadrature. This appears 
justified, since theoretical correlations in the calculations of the total 
branching fraction and the event fraction $F(E_0)$ are small. In (\ref{upshot}) 
we have combined all uncertainties in quadrature. A more conservative 
approach would be to add the errors linearly, in which case the error becomes 
${}_{-0.53}^{+0.49}$. Compared with the result
$\mbox{Br}(\bar B\to X_s\gamma)=(3.15\pm 0.23)\cdot 10^{-4}$ obtained in 
\cite{Misiak:2006ab}, our central value in (\ref{final}) 
is lower by about 5\% and, more 
importantly, the perturbative uncertainty is larger by almost a factor of 2. 
Both changes are a result of significant two-loop corrections encountered at 
the intermediate and soft scales, $\mu_i\sim\sqrt{m_b\Delta}$ and 
$\mu_0\sim\Delta$.

Our theoretical prediction for the $\bar B\to X_s\gamma$ decay rate is 
consistent with the present experimental world average, as is reflected in 
the ratio 
\begin{equation}\label{NPratio}
   \frac{\mbox{Br}(\bar B\to X_s\gamma)_{\rm exp}}%
        {\mbox{Br}(\bar B\to X_s\gamma)_{\rm th}}
   = 1.19\pm 0.09_{\rm exp}\pm 0.10_{\rm th} \,.
\end{equation}
However, whereas for a long time the experimental result used to be lower 
than the theoretical one, it is now about 1.4 standard deviations larger. 
Since in many extensions of the Standard Model the contributions from New 
Physics are expected to interfere constructively with the Standard Model 
$b\to s\gamma$ amplitude, the situation has changed from one where 
New Physics models were rather tightly constrained to one where there is now
room for speculation about how the central number in (\ref{NPratio}) could be 
explained in terms of loop contributions containing 
new heavy particles. Consider, e.g., the case of type-II 
two-Higgs-doublet models. Whereas in the past there used to be bounds on the 
charged-Higgs mass from $\bar B\to X_s\gamma$ on the order of 500\,GeV, the ratio (\ref{NPratio}) could now be explained with a New Physics contribution from a charged Higgs in just that mass range.

In summary, we have presented the first NNLO prediction for the 
$\bar B\to X_s\gamma$ branching fraction in which the effects of a 
photon-energy cut $E_\gamma\ge 1.6$\,GeV have been properly taken into account. 
Low-scale perturbative corrections 
lower the prediction for the branching ratio and introduce a significant 
theoretical uncertainty even at NNLO. Our result is 
about 1.4 standard deviations lower than the world-average experimental value. 
This re-opens the door for explorations of New 
Physics contributions to rare flavor-changing $B$-decay processes. 

\bigskip
{\em Acknowledgments:~}
The research of T.B.\ was supported by the Department of Energy under Grant 
DE-AC02-76CH03000. The research of M.N.\ was supported by the National Science 
Foundation under Grant PHY-0355005. Fermilab is operated by Universities 
Research Association Inc., under contract with the U.S.\ Department of Energy.
\vfil

\end{document}